\documentclass[aps,pra,reprint,amsmath,amssymb,superscriptaddress,showpacs,nobibnotes]{revtex4-1}
\usepackage{graphicx,SIunits}
\usepackage{bm}     % bold math
\usepackage{hyperref} % add hypertext capabilities
\usepackage{dsfont}    % allows \mathds{}
\usepackage{color}
\usepackage[normalem]{ulem}
\usepackage{lipsum}
\DeclareMathOperator{\Tr}{Tr}

\begin{document}

%\title{Experimental verification of different structure of different nonlocal quantum correlations under amplitude damping decoherence}

\title{Nonlocal quantum correlations under amplitude damping decoherence}
%\\Asymmetric behaviour of different quantum correlation in amplitude damping decoherence\\ \textcolor{blue}{Better title?}}

\author{Tanumoy Pramanik}
\email{tanu.pra99@gmail.com}
\affiliation{Center for Quantum Information, Korea Institute of Science and Technology (KIST), Seoul, 02792, Republic of Korea}

\author{Young-Wook Cho}
\affiliation{Center for Quantum Information, Korea Institute of Science and Technology (KIST), Seoul, 02792, Republic of Korea}

\author{Sang-Wook Han}
\affiliation{Center for Quantum Information, Korea Institute of Science and Technology (KIST), Seoul, 02792, Republic of Korea}

\author{Sang-Yun Lee}
\affiliation{Center for Quantum Information, Korea Institute of Science and Technology (KIST), Seoul, 02792, Republic of Korea}

\author{Sung Moon}
\affiliation{Center for Quantum Information, Korea Institute of Science and Technology (KIST), Seoul, 02792, Republic of Korea}
\affiliation{Division of Nano \& Information Technology, KIST School, Korea University of Science and Technology, Seoul 02792, Republic of Korea}

\author{Yong-Su Kim}
\email{yong-su.kim@kist.re.kr}
\affiliation{Center for Quantum Information, Korea Institute of Science and Technology (KIST), Seoul, 02792, Republic of Korea}
\affiliation{Division of Nano \& Information Technology, KIST School, Korea University of Science and Technology, Seoul 02792, Republic of Korea}

\date{\today} 

\begin{abstract}
\noindent 
Different nonlocal quantum correlations of entanglement, steering and Bell nonlocality are defined with the help of local hidden state (LHS) and local hidden variable (LHV) models. Considering their unique roles in quantum information processing, it is of importance to understand the individual nonlocal quantum correlation as well as their relationship. Here, we investigate the effects of amplitude damping decoherence on different nonlocal quantum correlations. In particular, we have theoretically and experimentally shown that the entanglement sudden death phenomenon is distinct from those of steering and Bell nonlocality. In our scenario, we found that all the initial states present sudden death of steering and Bell nonlocality, while only some of the states show entanglement sudden death. These results suggest that the environmental effect can be different for different nonlocal quantum correlations, and thus, it provides distinct operational interpretations of different quantum correlations. 
\end{abstract}

%\pacs{42.25.-p, 03.67.Ud, 03.67.-a, 42.50.Ex}

%42.50.Ex Quantum information, optical implementation
%42.50.Gy Phase coherence, quantum optics
\maketitle
\section{Introduction}

Nonlocal quantum correlations are not only significant due to their foundational aspects in quantum information theory, but also their applications in various quantum information processing tasks. According to the different local  models based on the properties of underlying systems, nonlocal quantum correlations can be categorized into three different forms of entanglement, EPR (Einstein-Podolsky-Rosen) steering, and Bell nonlocality~\cite{E_rev, B_Rev, Jones07_2}. A bipartite quantum system is entangled if it cannot be written as a statistical mixture of products of local states of individual systems. Therefore, for a bipartite entangled state, the correlation cannot be described by local hidden state (LHS)-LHS model. If we weaken the LHS-LHS model to LHS-local hidden variable (LHV) model, i.e., one of the systems is not trusted as a quantum system, then the non-separability becomes EPR steering~\cite{Jones07_1, Jones07_2}. If we further relax the condition to LHV-LHV model, then the non-separability defines Bell nonlocality~\cite{Bell, CHSH, Rev_BN}. Therefore, three forms of nonlocal quantum correlations are interconnected via their definitions. In particular, all Bell nonlocal states are steerable, and all steerable states are entangled. However, there exist some entangled states which are not steerable, and some steerable states are not Bell nonlocal. Therefore, we can explicitly present the relationship between nonlocal quantum correlations as, Bell nonlocality $\subset$ EPR steering $\subset$ Entanglement. 

In practice, nonlocal quantum correlations are used as resources of quantum information processing. Entanglement is known as a basic resource for many quantum information processing tasks such as quantum teleportation~\cite{Tele, Tele_Exp1, Tele_Exp2}, quantum communication~\cite{SDC, QKD1,QKD2, QKD3}, and quantum computation~\cite{QCom, Ent_Speed}. However, in order for entanglement to play roles, both systems should be trusted as quantum systems, and there should be no quantum hacking attempts to both systems. On the other hand, EPR steering and Bell nonlocality can play roles in the quantum information processing even when there exist quantum hacking attacks on one of the systems~\cite{1sDIQKD}, and both systems~\cite{QKD2, DIQKD_1, DIQKD_2, DIQKD_3, DIQKD_4}, respectively.

In real world implementation, quantum systems interact with the environment, and it usually causes unavoidable decoherence. As a result, quantum correlations usually gradually decrease with the increasing interaction time, and completely vanish after an infinite time of interaction~\cite{ESD_2, ESD_4, ESD_4_1,ESD_4_2}.  Remarkably, the system-environment interaction sometimes causes much faster degradation of quantum correlations, so the quantum system can completely lose quantum correlations in finite time of interaction. This phenomenon is known as the sudden death of quantum correlations~\cite{ESD_2, ESD_1, ESD_3, ESD_4, ESD_5, ESD_5_1}. We also note that the environmental interaction sometimes increases quantum correlations in certain circumstances~\cite{DE_1,DE_2,DE_3, DE_4}.

%As a result of decoherence, quantum correlation usually gradually decreases with the increasing interaction time, but, under special circumstance, quantum correlation, say, entanglement can be generated.  For example, when quantum systems interact with the common environment, entanglement between quantum systems can be created from separable states~\cite{DE_1,DE_2,DE_3, DE_4}. In the case of interaction of two-qubit with uncorrelated environment, quantum correlation completely vanish after an infinite time of interaction~\cite{ESD_2, ESD_4}. }

%%%%%%%%%%%%%%%%%%%%%%%%%
\begin{figure*}[t]
\includegraphics[width=7.0in]{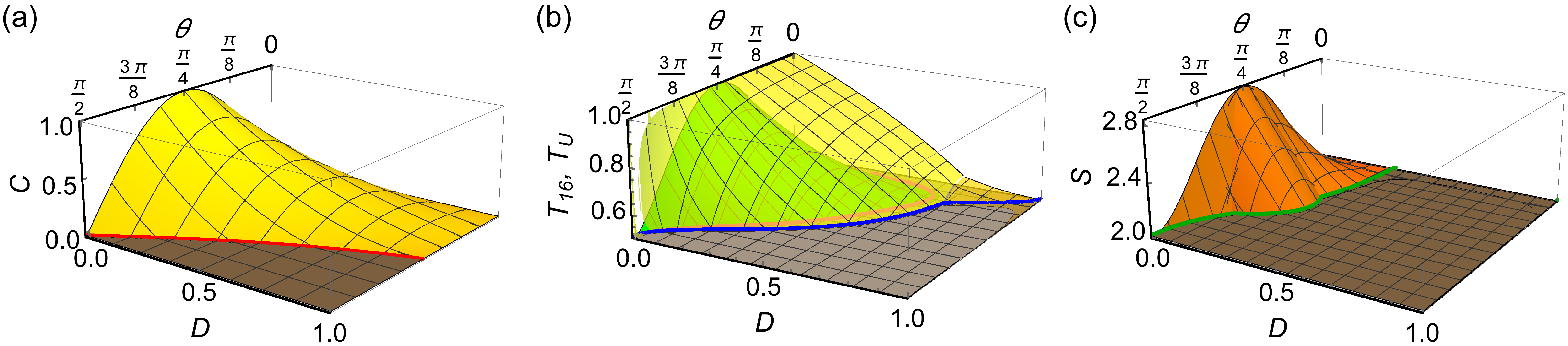}
\caption{(a) Concurrence $C(\theta,D)$, (b) EPR steering (green) and unsteering (yellow) parameters of $T_{16}(\theta,D)$ and $T_U(\theta,D)$, and (c) Bell parameter $S(\theta,D)$with respect to $\theta$. Above the $C=0$, $T_{16}=0.503, T_U=0.503$, and $S=2$ planes indicate the existence of nonlocal quantum correlations. 
, blue and green curves present the boundaries of sudden death phenomena. }
\label{3D}
\end{figure*}
%%%%%%%%%%%%%%%%%%%%%%%%%

It has been widely studied the effect of decoherence on entanglement both in theory and experiment~\cite{ESD_1, ESD_2, ESD_3, ESD_4, ESD_5}. However, there are only a few theoretical studies on other nonlocal quantum correlations~\cite{SSD,SSD_11,SSD_22, BNSD_1}. These studies deal with the entanglement sudden death (ESD)~\cite{ESD_1, ESD_2, ESD_3, ESD_4} and Bell nonlocality sudden death (BNSD)~\cite{BNSD_1}, however, the study of EPR steering sudden death (SSD) is still missing. Moreover, all of these works are limited to one of the nonlocal quantum correlations, and thus they fail to present unified results of the environmental effect on various nonlocal quantum correlations. Considering their relationship and unique roles in quantum information processing, it is of importance to investigate the dynamics of various nonlocal quantum correlations in the presence of decoherence.

In this paper, we theoretically and experimentally investigate entanglement, EPR steering, and Bell nonlocality under an amplitude damping channel (ADC). We found that different quantum correlations present very different environmental effects. For example, in our scenario, all the states present SSD and BNSD, while ESD happens for only some of the initial states. Moreover, we can prepare two different bipartite states with an equal amount of entanglement, but one of them shows sudden death of all nonlocal quantum correlations while the other only shows steering and Bell nonlocality sudden deaths, but not ESD. Therefore, in the presence ADC, entanglement behaves very differently from the other nonlocal quantum correlations, and it provides distinct operational interpretations of different nonlocal quantum correlations.

\section{Theory}
%\noindent
%{\bf Amplitude damping channel (ADC).} 
\subsection{Amplitude damping channel}
The interaction between the system $S$ and the environment $E$ via ADC  with the interaction strength of $0\le D\le1$ can be modeled as~\cite{lee11,kim12} %$|0\rangle_S|0\rangle_E  \rightarrow  |0\rangle_S|0\rangle_E$ and $1\rangle_S|0\rangle_E  \rightarrow  \sqrt{1-D}  |1\rangle_S|0\rangle_E  + \sqrt{D}  |0\rangle_S|0\rangle_E$.
%%%%%%%%%%%%%%%%%%
\begin{eqnarray}
|0\rangle_S|0\rangle_E & \rightarrow & |0\rangle_S|0\rangle_E, \nonumber \\
|1\rangle_S|0\rangle_E & \rightarrow & \sqrt{1-D}  |1\rangle_S|0\rangle_E  + \sqrt{D}  |0\rangle_S|0\rangle_E.
\label{ADC}
\end{eqnarray}
%%%%%%%%%%%%%%%%%%
Here, we assume that the environment is initially in  $|0\rangle_E$. Let us consider a two-qubit system is initially prepared in a pure state of $|\psi_\theta\rangle = \cos\theta |0\rangle_A|0\rangle_B + \sin\theta |1\rangle_A|1\rangle_B$,
%%%%%%%%%%%%%%%%%%
%\begin{eqnarray}
%|\psi_\theta\rangle = \cos\theta |0\rangle_A|0\rangle_B + \sin\theta |1\rangle_A|1\rangle_B,
%\label{State_i}
%\end{eqnarray}
%%%%%%%%%%%%%%%%%%
where $0\le\theta\le\pi/2$ is the biasing parameter. Assuming both qubits $A$ and $B$ are under ADC with an equal interaction strength of $D$, the state becomes~\cite{kim12}
%%%%%%%%%%%%%%%%%%
\begin{eqnarray}
\rho_{\theta}^{D} =
\begin{pmatrix}
\alpha_{11} & 0 & 0 &  \alpha_{14} \\
0 & \alpha_{21} & 0 & 0 \\
0 & 0 & \alpha_{21} & 0 \\
\alpha_{14} & 0 & 0 & \alpha_{44}
\end{pmatrix},
\label{State_f}
\end{eqnarray}
%%%%%%%%%%%%%%%%%%
where $\alpha_{11}=\cos^2\theta + D^2 \sin^2\theta$, $\alpha_{14}=(1-D) \cos\theta\sin\theta$, $\alpha_{21}=(1-D)D \sin^2\theta$, and $\alpha_{44}=(1-D)^2 \sin^2\theta$, respectively.

Now, we study entanglement, EPR steering, and Bell nonlocality of the state $\rho_{\theta}^D$. Here, we quantify the amount of entanglement with concurrence~\cite{Concurrence_1,Concurrence_2}. Bell nonlocality is determined by the Horodecki criterion which provides the necessary and sufficient condition for a $2\otimes 2$ dimensional system~\cite{Horo_Cri,Horo_Cri_2}. We apply the steering criterion developed in Ref.~\citep{Saunders, bennet12} to capture the steerability of a given state. Note that the steering criterion is necessary but not sufficient, and thus it cannot determine the unsteerability of a given state. In order to capture unsteerability, we employ the recently developed sufficient criterion of unsteerability~\cite{PE_St3}. Here, we only provide the results of the theoretical investigation. The detailed estimation procedures can be found in the supplementary materials.

%\noindent{\bf  Entanglement.} 
\subsection{Entanglement}

The concurrence of $\rho_{\theta}^D$ is given by
%%%%%%%%%%%%%%%%%%
\begin{eqnarray}
C(\theta,D) = \max\left[0,\,2 (1-D) \sin\theta  (\cos\theta - D \sin\theta)\right],
\label{C_f}
\end{eqnarray}
%%%%%%%%%%%%%%%%%%
and depicted in Fig.~\ref{3D}(a). All the initial states of $D=0$ has non-zero concurrence, and thus are entangled except for $\theta=0$ or $\pi/2$. As the interaction strength $D$ increases, concurrence decreases. One can find that entanglement vanishes, and the state $\rho_{\theta}^D$ becomes separable when $\cot\theta\leq D$. Therefore, the ESD occurs along the red line which corresponds to $D =  \cot\theta$.

Note that entanglement of the initial state, $C(\theta,D=0)=\sin2\theta$, is symmetrical with respect to $\theta=\pi/4$. Therefore, the initial states $|\psi_{\phi}\rangle$ and $|\psi_{\frac{\pi}{2}-\phi}\rangle$, where $0\le\phi<\frac{\pi}{4}$, have the same amount of entanglement.  This symmetry is broken as $C(\frac{\pi}{2}-\phi,D)<C(\phi,D)$ after the amplitude damping decoherence, $0<D$. This asymmetrical nature becomes more clear for the ESD, i.e., ESD occurs only for $|\psi_{\frac{\pi}{2}-\phi}\rangle$, and never happens for $|\psi_{\phi}\rangle$. It originates from the asymmetrical nature of the ADC where $|1\rangle$ experiences the damping decoherence while $|0\rangle$ is unaffected. 

We note that the non-zero concurrence provides the necessary and sufficient condition of the existence of entanglement in a two-qubit system~\cite{Concurrence_1, Concurrence_2}. Therefore, the entanglement sudden death described above is a real physical phenomenon although it has been investigated with the mathematical description of concurrence.

%\noindent{\bf  EPR steering.} 
\subsection{EPR steering}
LHS model restricts the correlation $P(a_\mathcal{A},b_{\mathcal{B}})$ between the measurement outcomes $a$ and $b$ of the observables $\mathcal{A}$ and $\mathcal{B}$ on the systems $A$ and $B$, respectively, as
%%%%%%%%%%%%%%%%%%%%
\begin{eqnarray}
P(a_{\mathcal{A}},b_{\mathcal{B}}) =  \sum_\lambda P(\lambda) P(a_{\mathcal{A}}|\lambda) P_Q(b_{\mathcal{B}}|\lambda),
\label{LHS}
\end{eqnarray}
%%%%%%%%%%%%%%%%%%%%
where $P(\lambda)$ is the distribution of hidden variables. The subscript $Q$ presents that Bob's probability distribution is obtained from the measurement of observable on the quantum system $B$. The joint probability distribution $P(a_{\mathcal{A}},b_{\mathcal{B}})$ for the shared bipartite state $\rho_\theta^D$ by Alice and Bob can be written as% $P_{\rho}(a_{\mathcal{A}},b_{\mathcal{B}})=\Tr\Big[\Big(\frac{I+(-1)^a \mathcal{A}}{2}\otimes\frac{I+(-1)^b \mathcal{B}}{2}\Big)\rho_\theta^D\Big]$.
%%%%%%%%%%%%%%%%%%%%
\begin{eqnarray}
P_{\rho}(a_{\mathcal{A}},b_{\mathcal{B}})=\Tr\Big[\Big(\frac{I+(-1)^a \mathcal{A}}{2}\otimes\frac{I+(-1)^b \mathcal{B}}{2}\Big)\rho_\theta^D\Big]
\label{JPD}
\end{eqnarray}
%%%%%%%%%%%%%%%%%%%%

The experimentally testable steering criterion can be derived with the help of the LHS model of Eq.~(\ref{LHS}). As quantum probability distribution $\big\{P_Q(b_{\mathcal{B}}|\lambda)\big\}$ for the measurement of non-commuting observables are bounded by the uncertainty principle, the correlation $\big\{P(a_{\mathcal{A}},b_{\mathcal{B}})\big\}$ is also bounded by the uncertainty principle. Several steering criteria have been derived based on different forms of uncertainty relation along with the LHS model~\cite{PE_St2,PE_St1, St_C1_2,St_C2,PE_St3}. 

Here, we employ the most widely accepted steering criterion of Ref.~\cite{Saunders, bennet12} as 
%$T_m=\frac{1}{m} \sum_{k=1}^m \langle \alpha_k  (\hat{n}_k\cdot\vec{\sigma}^B)\rangle \leq C_m$,
%%%%%%%%%%%%%%%%%%
\begin{eqnarray}
T_m=\frac{1}{m} \sum_{k=1}^m \langle \alpha_k  (\hat{n}_k\cdot\vec{\sigma}^B)\rangle \leq C_m,
\label{Steer_m}
\end{eqnarray}
%%%%%%%%%%%%%%%%%%
where $m$ is the number of the measurement settings of Alice and Bob, and the random variable $\alpha_k\in\{0,1\}$ is Alice's measurement result for $k$-th measurement.  Bob's $k$-th measurement corresponds to the spin measurement along the direction $\hat{n}_k$ and $\vec{\sigma}^B\in\{\sigma_x,\sigma_y,\sigma_z\}$, where $\sigma_x,\sigma_y,\sigma_z$ are the Pauli spin operators. $C_m$ is the maximum value of $T_m$ when Bob's system can be described by LHS model. The violation of Eq.~(\ref{Steer_m}) guarantees the steerability of the shared bipartite state $\rho_\theta^D$. The efficiency of the Eq.~(\ref{Steer_m}) increases with $m$, i.e., for a larger $m$, Eq.~(\ref{Steer_m}) captures larger set of steerable states. Here, we follow the technique used in the Refs.~\cite{Saunders, bennet12, Cavalcanti13} to increase the number of measurement settings, $m$. In Refs.~\cite{Saunders, bennet12, Cavalcanti13}, the vertices of the three-dimensional Platonic solids are used to design the measurement directions. There are only five three-dimensional Platonic solids with 4, 6, 8, 12, and 20 vertices. The measurement directions are chosen along the line by joining a vertex with its diametrically opposite vertex, except the Platonic solid with 4 vertices. With that, we can obtain $3,~4,~6,~10$ measurement settings from the Platonic solids with  6, 8, 12, and 20 vertices, respectively. We can increase the number of measurement settings by combining the measurement directions from the four Platonic solids. Here, we have chosen $m=16$ measurement settings 
by combining the axes of a dodecahedron (the Platonic solids with 20 vertices) and its dual, the icosahedron (the Platonic solids with 12 vertices). Note that we found that $m=16$ measurement settings capture larger sets of steerable states than other possible combinations using 4 Platonic solids in our scenario. In this case, steerability is guaranteed by the violation of the following inequality~\cite{bennet12,Saunders}.

%Note that, the steering criterion of Eq.~(\ref{Steer_m}) captures more steerable states with a larger number of measurement settings., i.e., its effectiveness to uncover steerable states increases with larger number of measurement settings. Therefore, it becomes tightest for $m\rightarrow\infty$~\cite{Saunders}. Here, we consider $m=16$, and $\hat{n}_k$ along the vertex-to-vertex axes of a geodesic solid which is created by combining the axes of dodecahedron and its dual, the icosahedron. In this case, steerability is guaranteed by the violation of the following inequalaity~\cite{bennet12,Saunders}
%%%%%%%%%%%%%%%%%%
\begin{eqnarray}
T_{16}(\theta,D)=\frac{1}{16} \sum_{k=1}^{16} \langle \alpha_k  (\hat{n}_k\cdot\vec{\sigma}^B)\rangle \leq  C_{16}=0.503.
\label{Steer_16}
\end{eqnarray}
%%%%%%%%%%%%%%%%%%

Since the steering criterion Eq.~(\ref{Steer_16}) is necessary, but not sufficient, it does not guarantee unsteerability. The unsteerability of the state $\rho_{\theta}^D$ can be verified with the help of the sufficient criterion of unsteerability derived in Ref.~\cite{PE_St3}. According to this criterion, the unsteerability of $\rho^D_{\theta}$ is determined when 
%%%%%%%%%%%%%%%%%%
%\begin{eqnarray}
%T_{U}(\theta,D)= 0.503\cdot \max\left[\alpha, \frac{2 \cos\theta \sqrt{1-D}}{\sqrt{\gamma}}\right] \leq 0.503,
%\label{Unsteer}
%\end{eqnarray}
%%%%%%%%%%%%%%%%%%
%%where $\gamma=\cos^2\theta+D\sin^2\theta$ and $\alpha=\{D^2 (\gamma - (1- D) \sin^2\theta)^2 +2 (1-D) \gamma\}/\gamma^2$, respectively. 
%$T'_{U}(\theta,D)= \max\left[\alpha, \frac{2 \cos\theta \sqrt{1-D}}{\sqrt{\gamma}}\right] \leq 1$,
%%%%%%%%%%%%%%%%%%
\begin{eqnarray}
t_{U}(\theta,D)= \max\left[\alpha, \frac{2 \cos\theta \sqrt{1-D}}{\sqrt{\gamma}}\right] \leq 1,
\label{Unsteer}
\end{eqnarray}
%%%%%%%%%%%%%%%%%%
where $\gamma=\cos^2\theta+D\sin^2\theta$ and $\alpha=\{D^2 (\gamma - (1- D) \sin^2\theta)^2 +2 (1-D) \gamma\}/\gamma^2$. 

Let us define the normalized unsteering parameter $T_U$ as
%%%%%%%%%%%%%%%%%%
\begin{eqnarray}
T_{U}(\theta,D)= 0.503\cdot t_{U}(\theta,D) \leq 0.503,
\label{Unsteer}
\end{eqnarray}
%%%%%%%%%%%%%%%%%%
in order to present the steering and unsteering criteria in the same figure, see Fig.~\ref{3D}(b). The green and yellow surfaces show $T_{16}(\theta,D) > 0.503$ and $T_U(\theta,D) > 0.503$, respectively. 
Therefore, the states $\rho_{\theta}^D$ lie on the green surface are steerable. Note that, similar to entanglement, the steering parameter $T_{16}(\theta,D)$ becomes asymmetrical with respect to $\theta=\pi/4$ after ADC. The states $\rho_{\theta}^D$ becomes unsteerable when $T_{U}(\theta,D)\leq0.503$. Therefore, the SSD occurs for $T_{U}(\theta,D)=0.503$, and it is presented by a blue curve in the Fig.~\ref{3D}(b). It is remarkable that SSD happens for all the initial states, unlike ESD.

%%%%%%%%%%%%%%%%%%%%%%%%%
\begin{figure}[b]
\includegraphics[width=3in]{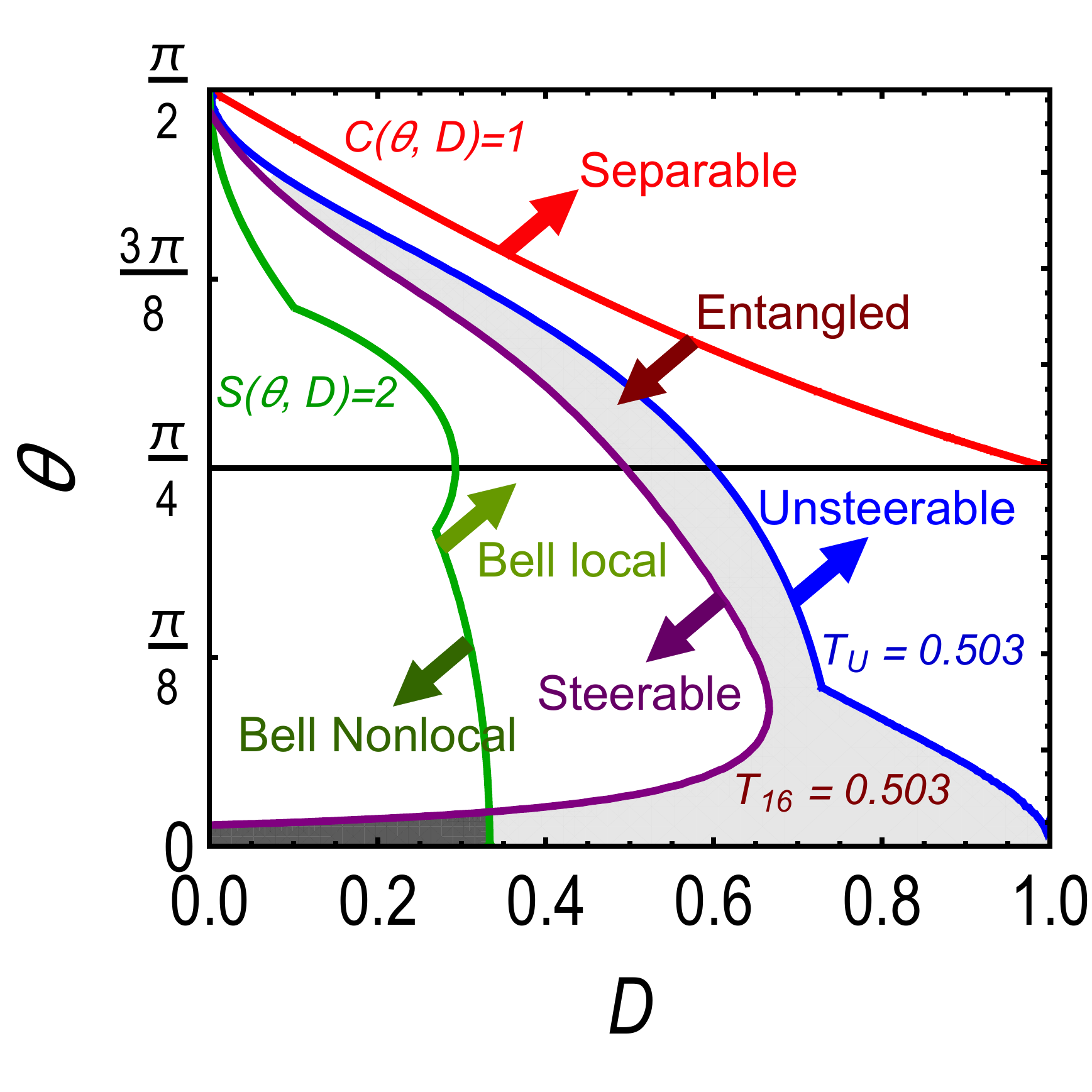}
\caption{The regions of various nonlocal quantum correlations for the bipartite state $\rho_{\theta}^D$. Red, purple, blue, and green lines correspond to $C(\theta,D)=0$, $T_{16}(\theta,D)=0.503$, $T_{U}(\theta,D)=0.503$, and $S(\theta, D)=2$, respectively.}
\label{Fig_Th}
\end{figure}
%%%%%%%%%%%%%%%%%%%%%%%%%

%\noindent{\bf  Bell nonlocality.} 
\subsection{Bell nonlocality}

The Bell nonlocality of a given state can be calculated from the correlation matrix $\lambda_{ij}^{\theta, D} = \Tr[\sigma_i\otimes\sigma_j\cdot\rho_{\theta}^D]$,
%%%%%%%%%%%%%%%%%%
%\begin{eqnarray}
%\lambda_{ij}^{\rho_{\theta}^D} = \Tr[\sigma_i\otimes\sigma_j\cdot\rho_{\theta}^D],
%\end{eqnarray}
%%%%%%%%%%%%%%%%%%
where $i,j\in\{x,y,z\}$~\cite{Horo_Cri,Horo_Cri_2}. The eigenvalues of $\left(\lambda_{ij}^{\theta, D}\right)^T\cdot \lambda_{ij}^{\theta, D}$, where the superscript $T$ denotes for transposition, are $\lambda_1=(\cos^2\theta + (1-2 D)^2\sin^2\theta)^2$, and $\lambda_2=(1-D)^2\sin^22\theta$ (with degeneracy). Therefore, the Bell parameter 
$S=\langle \alpha_1\beta_1\rangle + \langle \alpha_1\beta_2\rangle + \langle \alpha_2\beta_1\rangle -\langle \alpha_2\beta_2\rangle $, where $\{\alpha_1, \alpha_2\}$ and $\{\beta_1,\beta_2\}$ are sets of Pauli operators for Alice and Bob, respectively, 
is given by~\cite{Horo_Cri,Horo_Cri_2}
%%%%%%%%%%%%%%%%%%
\begin{eqnarray}
S(\theta, D) = \max\left[2\sqrt{2\lambda_2}, 2 \sqrt{\lambda_1+\lambda_2} \right].
\label{BI_rho}
\end{eqnarray}
%%%%%%%%%%%%%%%%%%
The state is Bell nonlocal if $S(\theta, D)>2$. The Bell parameter $S(\theta, D)$ is plotted in the Fig.~\ref{3D}(c). The orange curve shows the Bell nonlocality of the state $\rho_{\theta}^D$ and BNSD occurs  along the green line represented by $S(\theta,D) = 2$ where the orange surface touches the horizontal surface. Similar to SSD, BNSD occurs for all the initial states.

%\noindent{\bf Summary of the theoretical investigation.} 
\subsection{Sudden death of nonlocal quantum correlations}

The initial state $|\psi_{\theta}\rangle$ is entangled, steerable and Bell nonlocal for all values of $\theta$ chosen from the range of $0<\theta< \pi/2$. As a result of ADC, nonlocal quantum correlations decrease with the increasing interaction strength $D$. In order to compare the sudden death phenomena of various quantum correlations, we present the local-nonlocal boundaries of $C(\theta,D)=0$, $T_{16}(\theta,D)=0.503$, $T_U(\theta,D)=0.503$, and $S(\theta,D)=2$ in Fig.~\ref{Fig_Th}.

The red line corresponds to $C(\theta,D)=0$, and hence, it divides entangled states from separable states. It signifies that the state $|\psi_\theta\rangle$ with $0<\theta\leq \pi/4$ does not show ESD in ADC. The green curve presents $S(\theta,D)=2$, and thus show the BNSD boundary. It has discontinuities at $(\theta,D)\sim(0.35\pi,0.101)$ and $(0.21\pi,0.269)$ due to the maximization over two functions in Eq.~(\ref{BI_rho}). The purple and blue curves correspond to $T_{16}(\theta,D)=0.503$ and $T_U(\theta,D)=0.503$, and thus they are boundaries for steerable and unsteerable states, respectively. Between these two boundaries, there exists a undetermined area in gray where steerability of a given state cannot be concluded with the existing steering and unsteering criteria. As can be seen in the shaded by black region where the steering criterion fails to reveal the EPR steering for Bell nonlocal states, the steering criterion becomes invalid as $\theta\rightarrow0$. This non-ideal presentation can be improved by increasing the number of measurement settings~\cite{Saunders}.

It is interesting to compare the sudden death phenomena among various nonlocal quantum correlations. Although all quantum correlations of the initial state $|\psi_\theta\rangle$ are symmetrical with respect to the parameter $\theta$, they become asymmetrical after ADC. This happens due to the asymmetrical nature of ADC, i.e., ADC does not affect to $|0\rangle$ and $|1\rangle$ symmetrically. As discussed above, while both states $|\psi_\phi\rangle$ and $|\psi_{\phi+\pi/4}\rangle$, where $\phi < \pi/4$, have same amount of entanglement, ESD never happens for states $|\psi_\theta\rangle$. Whereas, all states with $0<\phi < \pi/2$ show SSD and BNSD. These results indicate that different nonlocal quantum correlations are affected by ADC in very different ways. 

% Therefore, the correlation defined solely by LHS model, e.g., entanglement can be protected in the presence of ADC. Whereas, sudden death of the correlations defined by LHV model, e.g., steering and Bell nonlocal cannot be suspended by the preparation method.

\section{Experiment}

%\noindent{\bf Experimental setup.} 
\subsection{Experimental setup}

%%%%%%%%%%%%%%%%%%%%%%%
\begin{figure}[b]
\includegraphics[width=3.4in]{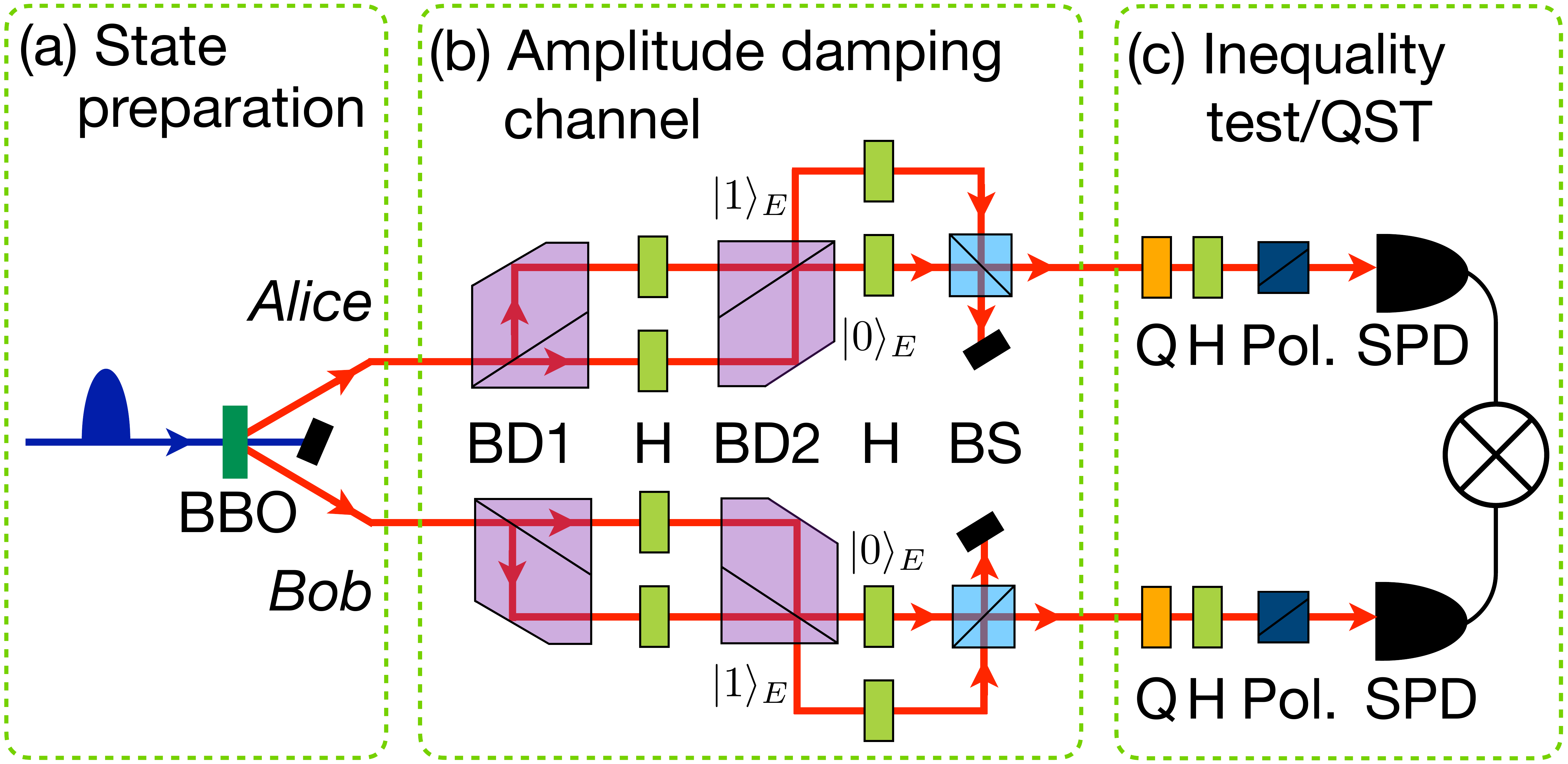}
\caption{ Experimental setup for (a) the initial state preparation, (b) amplitude damping channel, and (c) state measurement for inequality test and quantum state tomography. BD : beam displacer, H : half waveplate, Q : quarter waveplate, BS : beamsplitter, Pol. : Polarizer, SPD : Single-photon detector.}
\label{Setup}
\end{figure}
%%%%%%%%%%%%%%%%%%%%%%%

Figure~\ref{Setup} shows the experimental setup to explore nonlocal quantum correlations affected by ADC. The maximally entangled photon pair of $|\psi\rangle=\frac{1}{\sqrt{2}}(|00\rangle+|11\rangle)=\frac{1}{\sqrt{2}}(|HH\rangle+|VV\rangle)$ at 780~nm is generated at a sandwich BBO crystal via spontaneous parametric downconversion pumped by a femtosecond laser pulse. Here, $|H\rangle$ and $|V\rangle$ denote horizontal and vertical polarization states, respectively. The sandwich BBO crystal, which is composed of two type-II BBO crystals and a half waveplate in between, is specially designed for efficient generation of two-photon entangled states~\cite{wang16}.

In order to implement the amplitude damping channel (ADC), one needs to keep the probability amplitude of $|0\rangle$ unchanged while that of $|1\rangle$ changes to $|0\rangle$ with the probability $D$. Fig.~\ref{Setup}(b) shows our implementation of ADC with polarization qubits. Two beam displacers (BD) which transmit (reflect) horizontal (vertical) polarization state form a Mach-Zehnder interferometer. With the half waveplates (HWP, H) in the interferometer, one can independently control the ratio between two outputs $|0\rangle_E$ and $|1\rangle_E$ of BD2 for the horizontal and vertical polarization states. In the experiment, we set the HWP at the horizontal polarization path at $45^\circ$ in order to have all horizontal input photons at $|0\rangle_E$. On the other hand, the vertical polarization input state can be found both at $|0\rangle_E$ and $|1\rangle_E$ according to the angle of the HWP at the vertical polarization path. In order to cancel out the effect of the HWP in the interferometer, we position HWP at $45^\circ$ both at $|0\rangle_E$, and $|1\rangle$. The environment qubit is traced out by incoherently mixing $|0\rangle_E$ and $|1\rangle_E$ at a beamsplitter (BS)~\cite{lee11}.

%As shown in the Fig.~\ref{Setup}(c), two-qubit quantum state tomography (QST) and various inequality tests are conducted by two-qubit projective measurement and coincidence detection. In the experiment, concurrence $C$ and the unsteering parameter $T_U$ are calculated from the QST result whereas the Bell parameter $S$ and the steering parameter $T$ are directly obtained from the inequality test data. The experimentally implemented measurement settings for the steerability test can be found below. 

%%%%%%%%%%%%%%%%%%%%%%%

%\subsection{Quantum state tomography and measurements for Bell nonlocality and steerability}
As shown in the Fig.~\ref{Setup}(c), two-qubit quantum state tomography (QST) and various inequality tests are conducted by two-qubit projective measurement and coincidence detection. In the experiment, concurrence $C$ and the unsteering parameter $T_U$ are calculated from the QST result whereas the Bell parameter $S$ and the steering parameter $T_{16}$ are directly obtained from the inequality test data. The details of calculating entanglement and unsteerability as well as measurement settings for Bell nonlocaltity test and steering test can be found in the Appendices.

%To test Bell nonlocality of the prepared state, the possible sets of observables for the system $A$ are $\{\mathcal{A}_1=\sigma_x,\,\mathcal{A}_2=\sigma_y\}$ and $\{\mathcal{A}_1=\sigma_x,\,\mathcal{A}_2=\sigma_y\}$, and the corresponding the sets of measurement settings for the system $B$, $\{\mathcal{B}_1,\,\mathcal{B}_2\}$ has been derived in the Appendix~\ref{Apdx_Bell}. In the case of steering test, set of $16$ measurement settings corresponding to the spin along the direction of  vertex-to-vertex of dodecahedron and icosahedron, and the set of measurement settings for the system $A$ is derived in the Appendix~\ref{Apdx_St}.

%%%%%%%%%%%%%%%%%%%%%%%

\subsection{Experimental results}

For experimental verification of the effect of different quantum correlations in the presence of amplitude damping decoherence, we have prepared maximally entangled polarization photon pairs from spontaneous parametric down conversion. To test Bell nonlocality and steerability, we use CHSH  and steering inequalities derived in the Ref.~\cite{Saunders, bennet12}. To confirm unsteerability, we experimentally test the sufficient condition of unsteerability of Eq.~(\ref{Unsteer})  via quantum state~tomography~\cite{qst1,qst2}. The details of experiment can be found in the supplemental material.%~\cite{Supplement}.

%%%%%%%%%%%%%%%%%%%%%%%
\begin{figure}[t]
\includegraphics[width=3.4in]{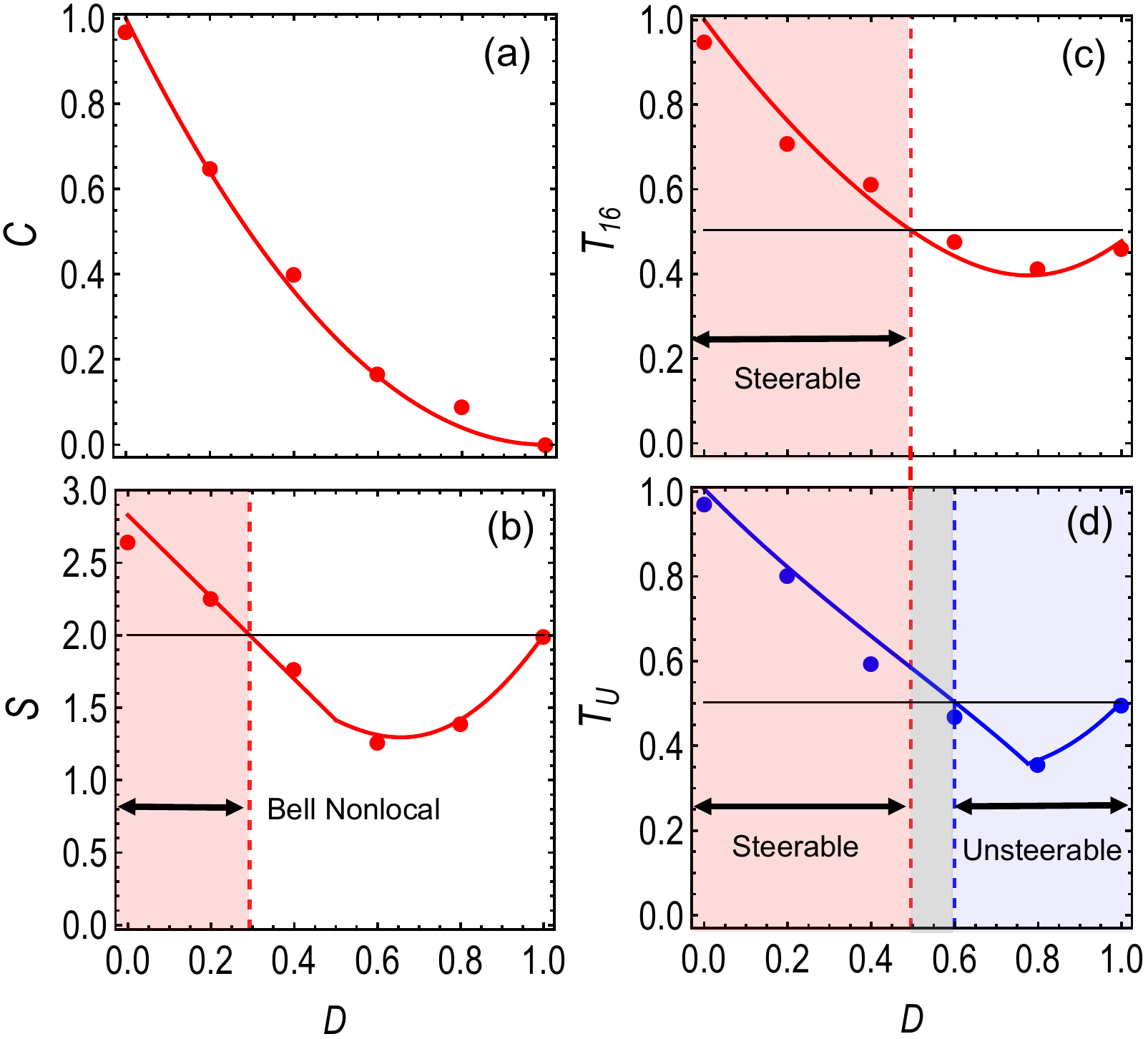}
\caption{Experimental results of the parameters of (a) entanglement, (b) Bell nonlocality, (c) EPR steering, and (d) unsteering for the initially maximally entangled state $\rho_{\theta=\pi/4}^D$, respectively. Red and blue lines and markers are theoretical and experimentally obtained values, respectively. Error bars are smaller than the size of makers. The horizontal black lines denote the local-nonlocal boundaries. Entanglement does not show the sudden death phenomenon, while Bell nonlocality and EPR steering sudden death happen. (d) The undetermined region for EPR steering is presented in gray.
}
\label{Fig_Exp}
\end{figure}
%%%%%%%%%%%%%%%%%%%%%%%

%For experimental verification of our theoretical analysis for the state $\rho_{\theta=\pi/4}^D$, we have performed experiments with polarization entangled photon pairs from spontaneous parametric down conversion. The Bell nonlocality and steerability are tested with the help of CHSH inequality and steering inequality with $16$ set of measurement settings on each sub-systems~\cite{bennet12,Saunders}. The concurrence and sufficient condition of unsteerability  are also tested via quantum state tomography~\cite{qst1,qst2}.

%\noindent{\bf Experimental results.} 
We present parameters of different nonlocal quantum correlations for the initially maximally entangled state $|\psi_{\theta=\pi/4}\rangle$ with respect to the interaction strength $D$ in Fig.~\ref{Fig_Exp}. Figure~\ref{Fig_Exp}(a) shows  theoretically and experimentally obtained concurrence $C$. It clearly shows that entanglement gradually degrades as $D$ increases, and the state becomes separable when $D=1$. Therefore, entanglement does not show sudden death phenomenon.

Fig.~\ref{Fig_Exp}(b) represents the Bell parameter $S$. The horizontal straight line corresponds to the upper bound of Bell inequality under LHV model, $S=2$. Similar to the concurrence, $S$ decreases as $D$ increases. More interestingly, $S$ becomes smaller than 2 even for $D<1$, that indicates sudden death of Bell nonlocality happens. In particular, we theoretically found that the sudden death of Bell nonlocality happens at $D\approx0.29$. Our experimental result coincides with the theoretical finding as the Bell nonlocal state at $D=0.2$ becomes Bell local at $D=0.4$. It is notable that unlike entanglement, Fig.~\ref{Fig_Exp}(b) shows the non-monotonous nature of  Bell local correlation (i.e., Bell parameter $S$ lies below 2) with respect to the decoherence parameter $D$. The values of $S$ decreases when $D$ increases from 0 to $0.66$, but for further increment of $D$ from $0.66$ to $1$, $S$ increases up to $2$. However, it never exceeds the the classical-quantum boundary of $S=2$. Due to the loss of quantum coherence measured by off-diagonal elements, different nonlocal quantum correlations, entanglement and Bell nonlocal correlation decrease gradually with the strength of decoherence and show monotonic behaviour. However, appearing and disappearing of the diagonal elements due to the effect of ADC is the source of non-monotonic behaviour of the local correlations, Bell local correlation (explained by local hidden variable theory).

The theoretical and experimental results of EPR steering and unsteerability are presented in Figs.~\ref{Fig_Exp}(c) and (d), respectively. The horizontal lines in the  Figs.~\ref{Fig_Exp}(c) and (d) are the upper bounds of steering inequality allowed by LHS model, i.e., $T_{16}=0.503$, and the upper bound of sufficient criterion of unsteerability, i.e., $T_U=0.503$, respectively. The vertical red (blue) line denotes the value of $D$ corresponding to the intersection between theoretical $T_{16}$ ($T_U$) and the horizontal line of $T_{16}=0.503$ ($T_{U}=0.503$). The light red shaded regions in both Figs.~\ref{Fig_Exp}(c) and (d) represent the range of $D$ for which the state $\rho_{\pi/4}^D$ is steerable. The light blue shaded region in Fig.~\ref{Fig_Exp}(d) shows the unsteerable region with respect to the parameter $D$. The steerable and unsteerable regions are separated by the gray region of $0.495\leq D\leq 0.6$ where the state cannot be concluded whether steerable or unsteerable with the existing criteria. The existence of unsteerable region verifies the EPR steering sudden death of the state $\rho_{\pi/4}^D$. Similar to the Bell local correlation, non-monotonic behaviour of unsteerability explained by local hidden state model occurs due to the effect of ADC on the diagonal elements.

\section{Conclusion}

We have theoretically and experimentally investigated different nonlocal quantum correlations of entanglement, EPR steering and Bell nonlocality under amplitude damping channel (ADC). Our results also show the dynamics of entanglement is completely different from those of EPR steering and Bell nonlocality in the presence of ADC. For example, in our scenario, entanglement sudden death depends on the preparation of initial entangled states whereas steering and Bell nonlocality sudden deaths happen for all the initial state. Therefore, our findings present clear theoretical and experimental evidences of structural difference between different nonlocal quantum correlations~\cite{tanu19}. They also indicate the operational difference of nonlocal quantum correlations in the presence of decoherence. Considering the fundamental and practical importance of nonlocal quantum correlations in quantum information science, our results not only provide better understanding, but also inspire various applications of quantum information. 

%{\color{red} Due to the effect of decoherence, different quantum correlations not only decrease with the strength of interaction but also sudden death of correlations occur. This effect of ADC on different correlations can be minimized, i.e., sudden death of correlations occur for higher strength of decoherence, with the help of the technique of weak measurement and its reversal~\cite{ADC_WK_1, kim12}.} 

\section*{Acknowledgement}

This work was supported by the KIST research program (2E29580) and the ICT R\&D program of MSIP/IITP (B0101-16-1355).

\onecolumngrid
\appendix 

\section{Calculation of entanglement}
\label{Apdx_Con}
%\subsection{Calculation of entanglement}

Entanglement of a bipartite state can be easily verified from its concurrence. If concurrence is positive, then the state is said to be entangled. The concurrence of the state $\rho_{\theta}^D$ can be calculated from the eigenvalues of $\Lambda^C=\rho_\theta^D\cdot\left(\sigma_y\otimes\sigma_y\cdot(\rho_{\theta}^D)^*\cdot\sigma_y\otimes\sigma_y\right)$, where the asterisk `$*$' stands for complex conjugation. For the state $\rho_{\theta}$, the eigenvalues of $\Lambda^C$ in decreasing order becomes
%%%%%%%%%%%%%%%%%%%%%%%%%%%%
\begin{eqnarray}
\lambda_1 &=& (1-D)^2\sin\theta^2\left(\sqrt{\cos^2\theta + D^2\sin^2\theta} + \cos\theta \right)^2, \nonumber \\
\lambda_2 &=& \lambda_3= (1-D)^2D^2\sin^4\theta, \nonumber \\
\lambda_4 &=& (1-D)^2\sin\theta^2\left(\sqrt{\cos^2\theta + D^2\sin^2\theta} - \cos\theta \right)^2.
\end{eqnarray}
%%%%%%%%%%%%%%%%%%%%%%%%%%%%
Using the above eigenvalues, the concurrence of the state $\rho_{\theta}^D$ can be calculated as
%%%%%%%%%%%%%%%%%%%%%%%%%%%%
\begin{eqnarray}
C(\theta,D) &=& \max\left[0,\,\sqrt{\lambda_1}-\sqrt{\lambda_2}-\sqrt{\lambda_3}-\sqrt{\lambda_3}\right] \nonumber \\
&=& 2 (1-D)\sin\theta(\cos\theta - D\sin\theta).
\end{eqnarray}

\section{Calculation of unsteerability}
\label{Apdx_Unst}

To derive sufficient criterion for an existing local hidden state (LHS) model of the state $\rho_{\theta}^D$, we need to transform it into  the canonical form $\varrho=\frac{1}{4}\left(\mathbb{I} + \vec{a}.\vec{\sigma} +\sum_{i=x,y,z} T_i\sigma_i\otimes \sigma_i\right)$,
where $\vec{a}\in\{a_x,a_y,a_z\}$ is Alice's local vector and $\{T_x,T_y,T_z\}$ forms a correlation matrix. $\rho_{\theta}^D$ can be converted to the above canonical form with the help of following transformation 
\begin{eqnarray}
\varrho_{\theta} = \mathbb{I}\otimes \left(  \left(\rho_{\theta}^D\right)^B  \right)^{-1/2} \cdot \rho_{\theta}^D \cdot \mathbb{I}\otimes\left(\left(\rho_{\theta}^D\right)^B\right)^{-1/2},
\end{eqnarray}
where $\left(\rho_{\theta}^D\right)^B=Tr_A[\rho_{\theta}]$. Then the sufficient criterion for unsteerability,
\begin{eqnarray}
T_U(\theta,D)=\max\left[a_z^2+ 2|T_z|, 2|T_x|\right] \leq 1
\end{eqnarray}
becomes
\begin{eqnarray}
\max\left[\alpha, \frac{2 \cos\theta \sqrt{1-D}}{\sqrt{\gamma}}\right] \leq 1,
\end{eqnarray}
where $\gamma=\cos^2\theta+D\sin^2\theta$ and $\alpha=\frac{D^2 (\gamma - (1- D) \sin^2\theta)^2 +2 (1-D) \gamma}{\gamma^2}$.

\section{Calculation of measurement settings for Bell nonlocality}
\label{Apdx_Bell}

Horodecki criterion provides maximum Bell violation of a given state in $2\otimes 2$ dimensional systems~\cite{Horo_Cri, Horo_Cri_2}. The measurement settings for both Alice and Bob corresponding to Bell violation as predicted by Horodecki criterion can be calculated with the help of Ref.~\cite{PE_BN_1,PE_BN_2,PE_BN_3}. To obtain Alice's and Bob's measurement settings corresponding to the Bell violation $S(\theta=\pi/4,\,D)$ of Eq.~(12) in the main text, let us consider two following scenarios. In the first scenario, Alice measures either observable $\mathcal{A}_1=\sigma_x$ or $\mathcal{A}_2=\sigma_y$ on her system $A$. Bob's choice of observables are
%%%%%%%%%%%%%%%%%%%%%%%%%%%%
\begin{eqnarray}
\mathcal{B}_1 &=&\sigma_x\cos\varphi_1 +\sigma_y\sin\varphi_1, \nonumber \\
\mathcal{B}_2&=&\sigma_x\cos\varphi_2 +\sigma_y\sin\varphi_2.
\label{BN_MS_rho}
\end{eqnarray}
%%%%%%%%%%%%%%%%%%%%%%%%%%%%
Then, the Bell parameter $S$ becomes
%%%%%%%%%%%%%%%%%%%%%%%%%%%%
\begin{eqnarray}
S_1(\theta=\pi/4,\,D) &=& (1-D)  \left(\cos\varphi_1 + \cos\varphi_2 - \sin\varphi_1 + \sin\varphi_2\right). 
\end{eqnarray}
%%%%%%%%%%%%%%%%%%%%%%%%%%%%
The maximum value of $S_1(\theta=\pi/4,\,D)$ can be found for $\varphi_1=7\pi/4$, and $\varphi_2=\pi/4$. Note that, $S_1(\theta=\pi/4,\,D)=S(\theta=\pi/4,\,D)$ for $0\leq D \leq 0.5$.

In the second scenario, Alice chooses observables from the set $\{\mathcal{A}_1=\sigma_x,\mathcal{A}_2=\sigma_z\}$ and Bob's set is given by
\begin{eqnarray}
\mathcal{B}_1 &=&\sigma_z\cos\chi_1 +\sigma_x\sin\chi_1, \nonumber \\
\mathcal{B}_2&=&\sigma_z\cos\chi_2 +\sigma_x\sin\chi_2.
\end{eqnarray}
In this case, the Bell parameter $S$ is given by
%%%%%%%%%%%%%%%%%%%%%%%%%%%%
\begin{eqnarray}
S_2(\theta=\pi/4,\,D) &=& (1-2(1-D)D)\cos\chi_1 - (1-2(1-D)D)\cos\chi_2 + (1-D)(\sin\chi_1+\sin\chi_2),
\end{eqnarray}
%%%%%%%%%%%%%%%%%%%%%%%%%%%%
which becomes maximum for $\chi_1=\arctan\left[(1-D)/(1-2(1-D)D)\right]$ and $\chi_2=\pi + \arctan\left[-(1-D)/(1-2(1-D)D)\right]$. In this scenario, $S_2(\theta=\pi/4,\,D)=S(\theta=\pi/4,\,D)$ for $0.5\leq D\leq 1$. Therefore, when decoherence parameter lies in the range $0\leq D\leq 0.5$, Alice and Bob choose the first scenario, otherwise, they choose the second scenario.

\section{Calculation of Measurement settings for steerability}
\label{Apdx_St}

In order to test the steering inequality with $16$ measurement settings on each subsystem, Bob chooses spin measurement along vertex-to-vertex of dodecahedron and icosahedron, Alice's measurement settings are calculated by maximizing $T_{16}$. Here, Bob's direction of $i$th spin measurement and Alice's direction of corresponding spin measurement settings are given by $\mathcal{B}_i\in\{n_x^i,n_y^i,n_z^i\}$ and $\mathcal{A}_i\in\{\sin\alpha_i \cos\beta_i,\sin\alpha_i \sin\beta_i, \cos\alpha_i\}$, respectively. The above measurement settings $\{\mathcal{A}_i,\mathcal{B}_i\}$ maximize the expectation value $\langle\mathcal{A}\mathcal{B}\rangle$ for the shared state $\rho_\theta^D$. $16$ set of measurement settings are given below
\begin{eqnarray}
\{\mathcal{A}_1,\mathcal{B}_1\} &\equiv & \{\{ \frac{1}{\sqrt{3}},\frac{1}{\sqrt{3}},\frac{1}{\sqrt{3}} \},\{\alpha_1=\arctan\left[-\frac{\gamma_1}{\delta_1}\right],\beta_1=\arctan[-1]   \}\}, \nonumber\\
\{\mathcal{A}_2,\mathcal{B}_2\} &\equiv & \{\{  - \frac{1}{\sqrt{3}},\frac{1}{\sqrt{3}},\frac{1}{\sqrt{3}}  \},\{  \alpha_2=\alpha_1 ,\beta_2=\frac{5\pi}{4} \}\}, \nonumber\\
\{\mathcal{A}_3,\mathcal{B}_3\} &\equiv & \{\{  \frac{1}{\sqrt{3}},-\frac{1}{\sqrt{3}},\frac{1}{\sqrt{3}}  \},\{  \alpha_3=\alpha_1 ,\beta_3=\frac{\pi}{4} \}\}, \nonumber\\
\{\mathcal{A}_4,\mathcal{B}_4\} &\equiv & \{\{  \frac{1}{\sqrt{3}},\frac{1}{\sqrt{3}},- \frac{1}{\sqrt{3}}  \},\{   \alpha_4=\pi + \arctan\left[-\frac{\gamma_1}{\delta_4}\right],\beta_4=-\frac{\pi}{4} \}\}, \nonumber\\
\{\mathcal{A}_5,\mathcal{B}_5\} &\equiv & \{\{  0,\frac{a}{b}, ab  \},\{  \alpha_5=\arctan\left[-\frac{\gamma_5}{\delta_5}\right], \beta_5=\frac{\pi}{2} \}\}, \nonumber\\
\{\mathcal{A}_6,\mathcal{B}_6\} &\equiv & \{\{ 0,-\frac{a}{b}, ab   \},\{  \alpha_6=\alpha_5, \beta_6=\frac{3\pi}{2}  \}\}, \nonumber\\
\{\mathcal{A}_7,\mathcal{B}_7\} &\equiv & \{\{  \frac{a}{b}, ab,0  \},\{  \alpha_7=\frac{\pi}{2},\beta_7=\arctan\left[-\frac{3+\sqrt{5}}{2}\right]  \}\}, \nonumber\\
\{\mathcal{A}_8,\mathcal{B}_8\} &\equiv & \{\{  -\frac{a}{b}, ab,0   \},\{  \alpha_8=\frac{\pi}{2},\beta_8=\pi+\arctan\left[\frac{3+\sqrt{5}}{2}\right]   \}\}, \nonumber\\
\{\mathcal{A}_9,\mathcal{B}_9\} &\equiv & \{\{  ab,0,\frac{a}{b}   \},\{  \alpha_9=\arctan\left[-\frac{\gamma_9}{\delta_4}\right],\beta_9=0  \}\}, \nonumber\\
\{\mathcal{A}_{10},\mathcal{B}_{10}\} &\equiv & \{\{  ab,0,-\frac{a}{b}  \},\{  \alpha_{10}=\pi + \arctan\left[\frac{\gamma_9}{\delta_4}\right], \beta_{10}=0  \}\}, \nonumber\\
\{\mathcal{A}_{11},\mathcal{B}_{11}\} &\equiv & \{\{ 0,  \frac{c}{d},  -\frac{1}{d} \},\{ \alpha_{11}=\pi+\arctan\left[\frac{\gamma_{11}}{\delta_{4}}\right],  \beta_{11}=\frac{3\pi}{2}   \}\}, \nonumber\\
\{\mathcal{A}_{12},\mathcal{B}_{12}\} &\equiv & \{\{  0,  \frac{c}{d},  \frac{1}{d}  \},\{  \alpha_{12}=\arctan\left[\frac{\gamma_{11}}{\delta_{4}}\right],  \beta_{12}=\frac{3\pi}{2}  \}\}, \nonumber\\
\{\mathcal{A}_{13},\mathcal{B}_{13}\} &\equiv & \{\{  \frac{c}{d},  \frac{1}{d}, 0  \},\{  \alpha_{13}= \frac{\pi}{2},  \beta_{13}=\arctan\left[-\frac{2}{1+\sqrt{5}}\right]  \}\}, \nonumber\\
\{\mathcal{A}_{14},\mathcal{B}_{14}\} &\equiv & \{\{  -\frac{c}{d},  \frac{1}{d}, 0  \},\{  \alpha_{14}= \frac{\pi}{2},  \beta_{14}=\pi + \arctan\left[\frac{2}{1+\sqrt{5}}\right]  \}\}, \nonumber\\
\{\mathcal{A}_{15},\mathcal{B}_{15}\} &\equiv & \{\{ \frac{1}{d}, 0, \frac{c}{d}   \},\{  \alpha_{15}=\arctan\left[-\frac{\gamma_{5}}{\delta_{15}}\right],  \beta_{15}=0  \}\}, \nonumber\\
\{\mathcal{A}_{16},\mathcal{B}_{16}\} &\equiv & \{\{ - \frac{1}{d}, 0, \frac{c}{d}   \},\{  \alpha_{16}=\alpha_{15},  \beta_{16}=\pi  \}\},
\end{eqnarray}
where
\begin{eqnarray}
\gamma_1 & =& \sqrt{2} (1-D) \sin2\theta, ~~~~ \delta_1= 4 D (1-D) \sin^2\theta - 1, \nonumber\\
\delta_4 &=& \cos^2\theta + (1-2D)^2 \sin^2\theta, \nonumber\\ % \gamma_4 & =& \gamma_1, ~~~~ 
\gamma_5 & =& 2 (1-D)\sin2\theta, ~~~~\delta_5=(3+\sqrt{5}) (2 D -1-2 D^2 - 2 D (1-D)\cos2\theta ),\nonumber\\
\gamma_9 & =& - (3+\sqrt{5}) (1-D) \sin\theta  \cos\theta, \nonumber\\ %~~~~\delta_9=\delta_4,
\gamma_{11} & =&  -  (1+\sqrt{5}) (1-D)  \sin\theta\cos\theta, \nonumber \\ % ~~~~ \delta_{11} = \delta_4,
\delta_{15} &=& (1+\sqrt{5}) (2 D -1-2 D^2 - 2 D (1-D)\cos2\theta ), \nonumber \\ %\gamma_{15} & =&  \gamma_5, ~~~~
a & =& c = \frac{1+\sqrt{5}}{2}, ~~~ b=\frac{1}{\sqrt{3}},~~~~ d=\sqrt{\frac{1}{2}\left(5+\sqrt{5}\right)}. \nonumber \\
\end{eqnarray}

\end{document}